# The Involution of Industrial Life Cycle on Atlantic City Gambling Industry

Jin Quan Zhou [(1)] Wen Jin He [(2)]

([1] Macau Polytechnic Institute, Macau, [(2)] Beijing Institute of Technology, Zhuhai, China)

**Abstract:** The industrial life cycle theory has proved to be helpful for describing the evolution of industries from birth to maturity. This paper is to highlight the historical evolution stage of Atlantic City's gambling industry in a structural framework covered by industrial market, industrial organization, industrial policies and innovation. Data mining was employed to obtain from local official documents, to verify the module of industrial life cycle in differential phases as introduction, development, maturity and decline. The trajectory of Atlantic City's gambling sector evolution reveals the process from the stages of introduction to decline via a set of variables describing structural properties of this industry such as product, market and organization of industry under a special industry environment in which industry recession as a result of theory of industry life cycle is a particular evidence be proved again. Innovation of the gambling industry presents the ongoing recovering process of the Atlantic City gambling industry enriches the theory of industrial life cycle in service sectors.

1. Introduction

The industry life cycle (ILC) theatrically describe an industry via four evolution phases on introduction, growth, maturity and decline, identifying by a number of variables such as market growth rate, potential demand growth, product variety, competitors, market share, entry barriers, technological innovation and guest behavior, in which any industry in a particular country or region could be presented as the four evolution stages (Saviotti and Pyka, 2008). Besides of qualitative description the ILC theory, an extensive literature on theory of ILC descripts an industrial evolution via imperial evidence to reveal the relationship between industrial structure, behavior and performance in different industry stages. Gort and Klepper (1982) develop the first ILC model (G-K) to present the ILC via changes of manufacturer's number as the industry evolution, which the theory gradually developed followed by the ILC theory of Agarwal (1998), and the ILC model (K-G) of Klepper (1997).

Classical literature mainly underlines the valid empirical evolution of ILC or industrial features in different industry and different market conditions. ILC theory on industrial evolution shed light on the industrial features via multi-indicators in terms of transformation of products and processes, changes in organization of firms, and other institute and government (Malerba and Orsenigo, 1996). The ILC concerns the changes in the number of manufactures through technological development, including product changes and economies of scale (Lomi et al., 2005). However, it targets only on the firm features including the legality and density of them, and the type of organization, especially in manufacturer areas. In addition, the ILC focus firms survive which reveals real changing from their entry and competence substitution on the company level (Winter, 2002). The evolution of ILC which identifies the development process on industry level affected under industrial innovation and industrial policy (Teece, 2007; Wang and Ahmed, 2007).

Practically, ILC empirical analysis exploits more than one indicator which presents the industrial features to reveal the industrial evolution need to cover the whole period of an industrial cycle.

The gambling industry can be generally defined as an industry in which individuals and institutions rely exclusively on gambling revenues from special customers named gamblers who enjoy gambling activities provided via gambling operators (Eadington, 1976) . Therefore, the gambling industry develops an economic activities categorizing to casinos, welfare and sports lottery, in which casino gambling and wagering are parts of entertainment and leisure industry that they have the general characteristics of service industry (Vogel, 2001). The service industry provides differential products which are non-physical, non-storage and synchronicity when serving and consuming.

Meanwhile, gambling products or services have the general characteristics of the service industry, as its special attraction to guests, which they have always been part of the tourism promotions to enhance tourism economy (Przybylski, 1998). Although the gambling industry was initially separated from the tourism industry by providing services mainly to adults, especially male adults, as results of competition mechanisms and market-oriented operations, which brings about an organic combination of gambling and tourism by gradually assembling hotels, restaurants, leisure and entertainments, resorts, conference and exhibition industries together (Ronald, 2013). More broadly, the gambling industry hosts closely to the tourism industry involving more economic and strategic issues (Eadington, 1984).

The industry and economy of gambling still lack interesting of mainstream economics researcher despite their rapidly growth in the global. Eadington (1999) state that the gambling economy has many distinct market characteristics in the particular aspects of products, prices and demand. As part of economic activity, government regulation exerts a significant influence on evolution of gambling industry (Braunlich, 1996; Gu, 2001), especially on historical urban planning and infrastructure development (Rubenstein, 1984). Furthermore, O'Donnell et al. (2012) find that industrial scale expansion is conducive to the development of gambling industry, but is greatly influenced by the external economic environment (Eadington, 2011), especially the external competition (Zheng, 2013). Hence, a research gap need aggregation on the development and evolution of the gambling industry.

The research arms of this paper are presented in four targets. (1) To verify the theory of ILC from an evidence of Atlantic City gambling industry. (2) To analyze the historical data of Atlantic City's gambling and depict the evolution procession of its ILC. (3) To identify the main factors leading the recession of Atlantic City's gambling industry on the perspectives of ILC. (4) To reveal how microeconomic factors in the evolution phase of ILC affects the dynamics of industry and its external the macroeconomic, market and demand characteristics.

## 2 Materials and Methods

The ILC theory provides a stylized description of the evolution of an industry, like product life cycle can describe its introduction, growth, mature and decline, to interpret the process of an industry evolution from small to large, from weak to strong (Storper, 1985). It highlights that any industrial structure, behavior and performance, as well as industrial policy issues, can be employed to analyze existing industrial structure and system are rationalized under given industrial base, historical culture and environment in each stage of regional industry evolution(Agarwal, 1998).

General speaking, the ILC theory propose a linking between the main industrial variables and the age of the industry, which each stage of the ILC has different characteristic variables in different industries (e.g. Agarwal, 1998). In the early stage of ILC, the rise of industry encourages the entering of a large number of firms and introducing of innovative products (Klepper, 1996). As a result, new enters with innovative product designs from outside of the industry are vying for market dominance (Gort and Klepper, 1982). In the development stage of ILC, the entry of new companies bring fierce competition resulting higher quality and lower prices of products followed by increased sales and the expansion of the industrial scale (Agarwal and Bayus, 2002). The maturity phase of ILC is marked by a shift of products process and changes of the number of films (Cohen and Klepper, 1996b; Klepper, 1996), while the interaction among institutional mechanisms, economic constraints and technological possibilities affects further expansion of the industrial scale that impedes the growth of products, market share, and the number of companies (Tushman and Murmann 1998; Murmann and Frenken 2006). During the decline stage of ILC, market share distribute to the most capable producers while sales would not grow indefinitely, which other companies withdrew from the industry, especially for those over-competition industries (Bertomeu, 2009), those small business and powerless innovators are exiting from industry (Klepper, 2002), then its recession happened (Klepper and Miller, 1995).

More broadly, the tourism ILC offer a theoretical reference for the gambling ILC because of an inextricably link between gambling industry and tourism industry (Przybylski, 1998). Organically combination of gambling industry with tourism industry, which makes gaming activity being a special attraction to tourist and leads a rapid widespread and development gambling boom around the world (Goodman, 1996). Concept of life cycle is systematically introduced into the area of tourism resort management, and developed a 5-stages tourism destination life cycle model while employing visit numbers as the characteristic variable on those stages (Butler, 1980). The model of the tourism ILC develops a dynamic and opened systematical framework (Haywood, 1992), influenced by some factors such as the decision-making of the tourism operators, the environmental factors of resorts, and the popularity, growth, accessibility and competition of destinations (Cooper and Jackson, 1989), as well as regulations of local government (Ioannides, 1992) where development and environmental policy are more important factors (Benedetto and Bojanic, 1993). The characteristic variables are mainly employed to analyze the life cycle in tourist destinations including the visit numbers, beds numbers, hotels, overnight stay, expenditure, tourism income, tax receipts, population, employment, tourism activities, and other tourism statistics (Berry, 2001).

Furthermore, industrial features in the stage of ILC are valid identified via more widely indicators such as number of enterprises, scale, market structure, prices and product quality (Kamat, 2010), combined with the characteristic variables of the tourist destination, as well as the features of the gambling industry, several variables can also be selected such as table game revenue, industry income, etc. (Moss, et al 2003, O'Donnell, 2012). In addition, considering some constraints and the feasibility of data collection, we determine the characteristic indicators that present life cycle phases of the gambling industry on five aspects: (1) Indicators for product's characteristics which present a basic judgment on life cycle evolution from product introduction to product replacement, as basic characteristics of evolution in ILC, where the quantity, price and quality of the products can reflect the industrial characteristics (Klepper, 1997; Jovanovic and MacDonald, 1994), especially reflect in gambling industry, the width and depth of the game products category, table game revenue, etc. (2) Indicators for industrial organizations which descript the evolution features of ILC by the number, size and income of companies (Murmann and Frenken, 2006; Eberle, et al., 2011), as well as the number, size and income of casino operators, employment. (3) Indicators for industrial market which indicates the industrial evolution by market demand and supply, including tourist market, stranded tourist (Butler, 2006; Murmann and Frenken, 2006), similar to changes of the number casinos and the scale of the gambling industry, which indicating via the gambling tourist market and stranded gamblers base on the gambling market demand and supply. (4) Indicators for industrial structure which presents the contribution of industry to regional economy by indicators of industry structure such as income ratio and taxation revenue ratio (Gort and Klepper, 1982; Mazzucato and Giovanni, 2002, Dinlersoz, and MacDonald, 2009), where the gambling industry contribute to the regional industry can be indicated by gambling market structure (rate of income and tax revenue). (5) Indicators for industrial policies which state industrial policy of government on industrial development (Tushman and Murmann, 1998; Murmann and Frenken, 2006; Potter and Watts, 2011), especially, the local government always regulate gambling industry by gambling licensing and tax policies.

The outlined literature provides the background for our empirical analysis and allows us select the characteristic indicators from the gambling ILC to examine the evolution of the gambling industry in Atlantic City. Due to data restrictions, the data on which we employed in our analysis are presented in Table 1. Data source abstracts from Annual Report 1978-2017, New Jersey Casino Control Commission.

**Table 1. Data description**

| Year | Casinos | Slot Machines | Tables | Employees | Table Revenue | Slot Revenue | Online Revenue | Visitors | Total Revenue |
|---|---|---|---|---|---|---|---|---|---|
| 1978 | 1 | 1,390 | 84 | 3300 | 75,491 | 58,582 |  | 7,008 | 134,073 |
| 1979 | 3 | 3,500 |  | 10100 | $188,135 | $137,345 | – | 9,465 | $325,480 |
| 1980 | 6 | 7,481 | 649 | 16700 | $374,846 | $267,827 | – | 13,822 | $642,673 |
| 1981 | 9 | 11,365 | 952 | 27600 | $629,525 | $470,256 | – | 19,084 | $1,099,781 |
| 1982 | 9 | 12,061 | 941 | 29100 | $799,174 | $693,989 | – | 22,955 | $1,493,163 |
| 1983 | 9 | 11,890 | 871 | 30,197 | $893,008 | $877,935 | – | 26,361 | $1,770,943 |
| 1984 | 10 | 14,341 | 1,039 | 33,681 | $971,669 | $980,099 | – | 28,466 | $1,951,768 |
| 1985 | 11 | 16,183 | 1,165 | 37,004 | $1,050,909 | $1,087,742 | – | 29,326 | $2,138,651 |
| 1986 | 11 | 16,525 | 1,198 | 37,262 | $1,095,843 | $1,185,361 | – | 29,932 | $2,281,204 |
| 1987 | 12 | 18,616 | 1,312 | 39,351 | $1,169,583 | $1,326,091 | – | 31,845 | $2,495,674 |
| 1988 | 12 | 19,534 | 1,323 | 42,134 | $1,241,196 | $1,493,577 | – | 33,138 | $2,734,773 |
| 1989 | 12 | 18,202 | 1,196 | 41,627 | $1,229,801 | $1,577,216 | – | 32,002 | $2,807,017 |
| 1990 | 12 | 21186 | 1355 | 45,241 | $1,227,258 | $1,724,323 | – | 31,813 | $2,951,581 |
| 1991 | 12 | 21654 | 1276 | 43,910 | $1,140,491 | $1,851,068 | – | 30,788 | $2,991,559 |
| 1992 | 12 | 22774 | 1144 | 44,240 | $1,102,168 | $2,113,802 | – | 30,705 | $3,215,970 |
| 1993 | 12 | 24561 | 1288 | 44,111 | $1,086,729 | $2,214,631 | – | 30,225 | $3,301,360 |
| 1994 | 12 | 27041 | 1310 | 44,894 | $1,125,388 | $2,297,146 | – | 31,321 | $3,422,534 |
| 1995 | 12 | 29057 | 1301 | 47,286 | $1,174,857 | $2,572,721 | – | 33,272 | $3,747,578 |
| 1996 | 12 | 32786 | 1439 | 48,956 | $1,187,631 | $2,626,022 | – | 34,042 | $3,813,653 |

| 1997 | 12 | 35057 | 1473 | 49,123 | $1,185,992 | $2,720,146 | – | 34,070 | $3,906,138 |
| --- | --- | --- | --- | --- | --- | --- | --- | --- | --- |
| 1998 | 12 | 36008 | 1373 | 48,492 | $1,207,839 | $2,825,158 | – | 34,300 | $4,032,997 |
| 1999 | 12 | 34718 | 1321 | 47,366 | $1,208,312 | $2,955,885 | – | 33,652 | $4,164,197 |
| 2000 | 12 | 36278 | 1306 | 47,426 | $1,212,748 | $3,087,584 | – | 33,184 | $4,300,332 |
| 2001 | 12 | 37483 | 1247 | 45,592 | $1,161,805 | $3,141,272 | – | 32,423 | $4,303,077 |
| 2002 | 12 | 38117 | 1177 | 44,820 | $1,119,890 | $3,261,516 | – | 33,188 | $4,381,406 |
| 2003 | 12 | 42378 | 1386 | 46,159 | $1,161,057 | $3,327,277 | – | 32,224 | $4,488,334 |
| 2004 | 12 | 41605 | 1443 | 45,501 | $1,250,285 | $3,556,412 | – | 33,230 | $4,806,697 |
| 2005 | 12 | 41231 | 1606 | 44,452 | $1,344,317 | $3,673,959 | – | 34,924 | $5,018,276 |
| 2006 | 12 | 35848 | 1658 | 42,545 | $1,413,998 | $3,803,615 | – | 34,534 | $5,217,613 |
| 2007 | 11 | 35615 | 1615 | 40,788 | $1,456,316 | $3,464,470 | – | 33,300 | $4,920,786 |
| 2008 | 11 | 34123 | 1628 | 38,585 | $1,412,460 | $3,132,501 | – | 31,813 | $4,544,961 |
| 2009 | 11 | 30782 | 1602 | 36,377 | $1,221,397 | $2,721,774 | – | 30,381 | $3,943,171 |
| 2010 | 11 | 28113 | 1573 | 34,145 | $1,087,696 | $2,477,350 | – | 29,328 | $3,565,046 |
| 2011 | 11 | 27048 | 1579 | 33,093 | $974,783 | $2,342,936 | – | 28,452 | $3,317,719 |
| 2012 | 12 | 26881 | 1630 | 32,823 | $810,776 | $2,183,960 | – | 27,227 | $2,994,736 |
| 2013 | 12 | 25306 | 1567 | 32,427 | $798,243 | $2,063,826 | $8,369 | 23,370 | $2,870,438 |
| 2014 | 8 | 17232 | 1174 | 24843 | $708,218 | $1,874,715 | $122,877 | 22,950 | $2,705,810 |
| 2015 | 8 | 16770 | 1170 | 23615 | $682,781 | $1,731,456 | $148,880 | 22,437 | $2,563,117 |
| 2016 | 7 | 14125 | 1046 | 22005 | $769,373 | $1,712,277 | $196,709 | 24,460 | $2,602,721 |
| 2017 | 7 | 14052 | 1032 | 22178 | $769,374 | $1,719,672 | $245,606 | 22,210 | $2,659,013 |

**3 Emergences and Development of Contemporary Gambling Industry in Atlantic City**

Atlantic City was once the largest tourist city in New Jersey known as the "World's Playground" as a tourist attraction. In the late 1960, many famous resorts continued to have low occupancy rates, most of them closed, turning to cheap apartments or nursing homes. Many of these hotels were demolished in the period leading up to the emergence of legal gambling. Exploring legal gambling has become an important step since the city was revitalized (Johnson, 2010). Atlantic City was legalized gambling in 1977 except Nevada. Resorts International opened its first casino on May 26, 1978, two years after a series of major casino operators, including Caesars, Bally's and Harrah's moved to Atlantic City. In 1988, Atlantic City surpassed Las Vegas as the largest gambling city in the United States, which gambling revenue peaked at $5.2 billion in 2006. Since then it has fallen year after year to $2.56 billion in 2015, a decline of more than half (Cooper, 2007). Five of the city's 12 casinos closed between 2013 and 2016 companying 20 percent of their jobs lost from 2014 with an unemployment rate of 6.3 per cent overpassing national level (4 per cent), Atlantic City once filed for bankruptcy protection in 2015. New Jersey's online gambling industry and sports betting opened up a differentiated market in 2013, with a growing 32.1 per cent to $197 million in 2016 that makeup the losses in the casino gambling and generating small profits while Casino gambling revenue has been reduced by 1% each year. Total gambling revenue was $2.6 billion in 2016 adding 1.5 percent from $2.56 billion in 2015, and the gambling industry's recession was stopped. Atlantic City's gambling industry has experienced ups and downs cycle within 40 years. After decades declining, Atlantic City tentative develop more diversifying industry to rescue Atlantic City's gambling industry in the depressed process , until

2012, sports gambling and the legal online gambling bring new opportunities to the Atlantic City gambling industry.

**Introduction phase of Atlantic City's gambling industry**. New Jersey has a long history of gambling, and the state has historically indulged in gambling more than any other state until it was banned in 1844. New Jersey issued lottery affording military bill during the Wars in France and India and the American Revolution, funding to establish Queen's College (now Rutgers University) and New Jersey College (now Princeton University) (Richard, 1988). In addition, Freehold Raceway,  an informal horse racing and the oldest racecourse in the United States, was hosted in 1830, and the Monmouth County Agricultural Association set annual fair and horse racing in 1854 (Jerry, 1992; Barbara, 2003). In fact, all gambling activities in New Jersey were theoretically illegal from 1894 to 1939, these laws however, were not fully enforced resulting widespread of book-taking, digital games, and slot machines. Many churches and other non-profit organizations openly run Bingos, and Freehold Raceway racehorses uninterrupted until 1939 till racecourse gambling was re-legalized (Barbara Pepe, 2003).

**Development phase of Atlantic City's gambling industry.** Atlantic City began legalize casinos in 1976 though voters in New Jersey voted against the legalization of casinos in 1974. For the case of economy rescuing, New Jersey become the second region with a legal casino in the United States since 1978. Resorts International (later known as The Resort of Merv Griffin; Atlantic City Casino Resort) opened the first legal casino in the eastern United States on May 26, 1978. Subsequently, Caesar's Boardwalk Regency (Atlantic City of Caesar) opened on June 26, 1979, followed by Bally's Park Place (Atlantic City of Barley) on December 29, 1979, The Brighton (Sands Hotel and Casino) on August 11, 1980, Del Webb's Claridge's and Hi-Ho Casino (Bally's Park Plaain on November 23, 1980. Harrah's Casino Hotel on November 23, 1980, Nugget (Atlantic City Hilton Casino Resort, ACH Atlantic Club Casino Hotel) on December 12, 1980, Playboy Hotel and Casino Trump's World Fair in April 14, 1981 and Golden Hce and Claridge Hotel on July 20, 1981 and Tropicana Casino and Resort Tropicana on November 26, 1981. With gambling industry booming, tourists are once flocking to Atlantic City. Despite of the surge in gambling revenues, some studies revealed that legalized casino gambling has not stopped crime and poverty increasing in Atlantic cities. However, Gambling tourism rose sharply during this period.

**Growth phase of Atlantic City's gambling industry.** Although Atlantic City legal gambling later than Las Vegas, it has become the top gambling resort in the United States, which bring its gambling industry a bright future within a short period. After entering the mid-1980, Harrah's Boardwalk Hotel Casino (later known as the Trump Plaza Hotel Casino opened on May 26, 1984, Trump Castle (later known as the Trump Marina Hotel and Casino) opened on June 19, 1985, Showboat Casino Hotel opened on April 3, 1987, and Trump Taj Mahal Casino Resort opened on April 2, 1990. Donald Trump became Atlantic City's gambling king after three casinos acquisitions. Trump held three license plates for 12 casinos in Atlantic City - Trump Castle, Trump Plaza, and Trump Taj Mahal (with an investment of nearly 1 billion to build, the tallest building and largest casino in New Jersey) by 1990. The Playboy Hotel and Casino was his the fourth casino in Atlantic City. Atlantic City gambling industry had revenues of 2.73 billion in 1988, becoming the largest casino city in the United States when gambling revenue on the Las Vegas Strip reached just 1,944 million in that year. Gambling was booming during this period.

**Maturity phase of Atlantic City's gambling industry**. Casino revenues in New Jersey have risen steadily since 1990, gambling industry revenue changed from 2.95 billion in 1990 to 4 billion in 1998, rechecked 4.2 billion in 2000 and 4.38 billion in 2002 within 12 casinos, which nearly 75 percent of gambling revenue come from slot machines. Trump Taj Mahal Casino Resort make biggest beneficiary from its hosted more than 4,700 slot machines. Whilst Atlantic City's rivals seem keen to aggregate as many slot machines as possible in the casino, competition has intensified among them. Until Wynn and Boyd's Borgata Hotel Casino and Spa opened on July 2, 2003, the true Las Vegas-style hotel and casino appeared in Atlantic City with 200 tables within 34 table poker rooms, in particular, casinos offer chic nightlife and restaurant services that Atlantic cities have never seen before. Poker game expands rapidly in Atlantic City while Borgata hosted series of televised World Poker Tours. Borgata is a game innovator to expand poker to 85 tables and positioning itself as the first poker destination on the East Coast. But Trump Taj Mahal Casino Resort remains the city's revenue leader, where Atlantic City's gambling revenue peaked at 5.2 billion in 2006. Gambling was stable during this period.

**Decline phase of Atlantic City's gambling industry.** Affected by the economic crisis and other factors, especially the rapid growth of casinos in areas around New Jersey, such as Pennsylvania has trigger intensive competition to attract customers. Atlantic City's gambling industry was surfing severe recession with Sands shut down in 2006. The revenues of remaining 11 casinos in Atlantic City fall by 4.8 million in 2007, followed by 4.5 billion in 2008 and 4 billion in 2009. Atlantic City's gambling industry keep decline to after 2010, Borgata became Atlantic City's industry leader and the only bright spot, but other less competitive casinos continually closing, and a one third of casinos in Atlantic City closed in 2014. Moreover, closing tide followed by Bally's Grand, Atlantic City Hilton and Showboat, Atlantic Club Casino Hotel. Revel Entertainment which opened on May 25, 2012 with a 2.4 billion investment obliged to close in September 2014. Only 8 casinos remain in Atlantic City keeping operation, including Bally's Atlantic City, Borgata Hotel Casino and Spa, Caesars Atlantic City, Golden Nugget Hotel Casino, Harrah's Resort Atlantic City, Resorts Casino Hotel and Tropicana. Casino and Resorts. As a result of gambling industry recession, a large number unemployed casino staff left Atlantic City.

**Recovery period of Atlantic City's gambling industry.** After the great recession in gambling, Atlantic City explored changes and found new recovery for its gambling industry. One result is that the Sports Betting Act is voted in January 2012 (Sarah, 2012). The law allows the state's 12 casinos and four racecourses to operate professional sports betting, but prohibits betting on races in New Jersey and games related to the New Jersey College team (Stacy, 2012). But it filed federal lawsuits against New Jersey for sports betting by NCAA, NBA, NFL and Hockey Leagues that prohibited in all states under the Professional and Amateur Sports Protection Act of 1992. New Jersey pushed ahead with a legalized sports betting program on May 14, 2018, as a result, legalized sports betting on June 11, 2018.

The Lesniak Act of New Jersey allowed online gambling for New Jersey residents above 21 year old in January 2011. Online gambling allowed at Atlantic City casinos which legislate designate computer servers that operate online gambling sites to be located at licensed Atlantic City casinos, while it are prohibited businesses in other region besides Atlantic City casinos (John, 2012). The Internet Gambling Act authorized casinos to operate online gambling for a 10-year probationary period, limited to 11 casinos in Atlantic City, and imposed a 15 percent tax on online gambling revenue instead of an 8 percent casino revenue tax (Ryan, 2013). The bill requires gamblers to be at least 21 years old and bet on computers within New Jersey. The gambler's location could be verified by the Global Positioning System (GPS), which allows future interstate agreements to authorize online gambling between multiple continents. 12 online casino operators offered hundreds of games, including games transforming from traditional casinos after November 21, 2013.

**4 Atlantic City gambling industry dynamic：Evolution and Analysis**

The gambling industry which has the evolution mechanism like other industries within its industrial cycle, together with its exclusive feature which the characteristic indicators are determined for describing the gambling life cycle such as industry revenue, table game income and product quantity, etc. Then, the data presented in table 1 ware employed to depict the evolution process of the life cycle of Atlantic City's gambling industry from multi-perspectives of gambling products, industrial organization, industrial market, industrial structure and industrial policy.

**4.1 Industry market and the evolution process of Atlantic City's gambling industry**

The total number of visitors to Atlantic City grew rapidly from 1978 to 1985, reached 7.08 million to 2.9326 million, following 34.924 million in 2005. The gambling industry continually recessed at a low market performance with 22.437 million tourists in 2015 since 2006. Atlantic City's gambling industry however emerges with a tremendous growth from 1978 to 1985, which total gambling revenue rise at a rate of 55.07 per cent average annual. There is a steady growth period from 1986 to 2006 with total gambling revenue rising at a rate of 4.37 per cent average annual, the growth curve turns from peak point at 2006 and begin continually falling until 2015.

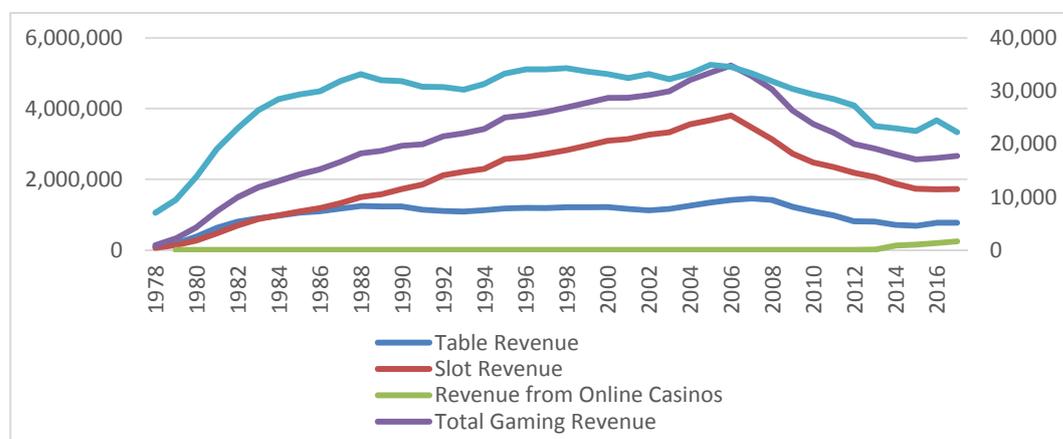

Figure 1 the dynamics of Industry life based on industry market.
（Resource：Annual report 1990-2017, New Jersey Casino Control Commission,
https://www.nj.gov/casinos/about/reports/）

Slot Revenue become the biggest scale gaming product when it surpassed table revenue in Atlantic City's gambling industry since 1983. It continued to grow at peak in 2006 and keep to decline since 2007 at an average -13.11 percent each year from 2007 and 2015. Table game revenues have presented stable feature since 1983 and have been falling since 2006. The average share of table income was 36.42 per cent from 1979 to 2014. Tables accounted for 58.33 per cent of total gambling revenue when slot machines share 41.67% of total revenue in 1980. Slot revenue exceeded that of table games after 1984 in Atlantic City's casinos while the average revenue share of slot machines was 63.45% of total revenue during the period 1979-2017. Slot machine has lost some of its share for online gambling which has continued to grow since it was launched in 2013. As a result, online gambling revenue increased with a market share from 0.29% in 2013 to 9.23% in 2017.

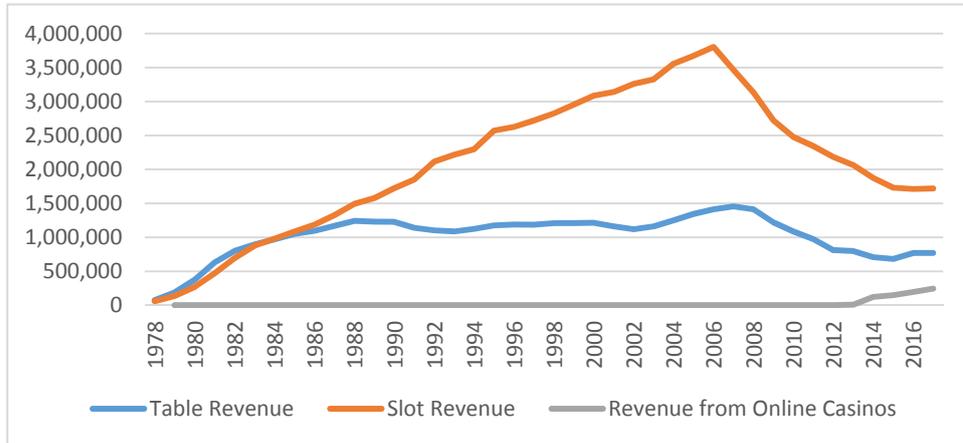

Figure 2 the dynamics of Industry life based on industry revenue.

（Resource：Annual report 1990-2017, New Jersey Casino Control Commission, https://www.nj.gov/casinos/about/reports/）

**4.2 Industrial organization and the evolution process of Atlantic City's gambling industry**

Atlantic City casino operators quickly added to 9 from 1978 to 1983 and reached the highest number of 12 in 1986. But it continued to close after 2013 and only 7 of 12 survived in 2017. The number of slot machines has continued to grow from 1978 to 2006 and begin sliding since after 2006. The number of tables has been stable with a little fluctuating between 1983 and 2005 and dropping rapidly between 2005 and 2013 since the industry declining. Atlantic City casino hair 49,123 employees in 1997, but it shrinks to 30 years ago where 22,178 employees are served in 2017 since gambling continued to fall from 2006.

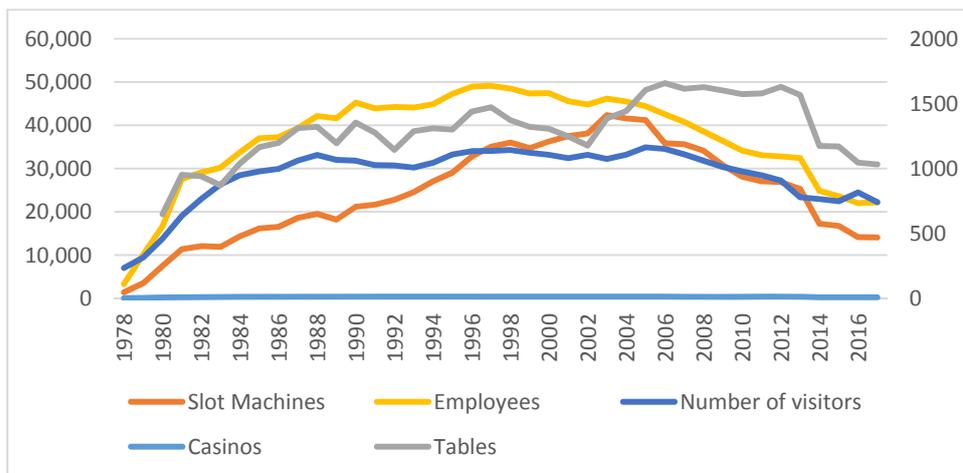

Figure 3 the dynamics of Industry life based on industry organization.

（Resource：Annual report 1990-2017, New Jersey Casino Control Commission, https://www.nj.gov/casinos/about/reports/）

**4.3 Gaming products and the evolution process of Atlantic City's gambling industry**

The number of low denomination slot machines between $0.10 and $0.20 has been increasing since 2005, while the number of denomination slot machines $0.25 and $0.50 has been declining steadily since 2001 in casinos at Atlantic City. In addition, table game category have evolved in a increasing variety in Atlantic City including Blackjack, Craps, Roulette, Spanish 21, Three Card Poker, Baccarat, Mini Baccarat, Big Six Keno, Let It Ride Poker, Pai Gow, Pai Gowker, Po Four Poker, Sic Bo, Caribbean Stud Poker, Casino War, Double Attack, Blackjack Texas, Hold'em Bonus Poker, Flop Poker, Ultimate Texas, Hold'em Asia Poker, Mississippi Stud, Criss Cross Poker, High Card Flush, Head Up Hold'em, Double Draw Poker, Pack's Poker, Football Kings, and Electronic Table Games. Atlantic City's casinos holds 12 tapes table games in 2001, despite of a decline in table revenues from 2014 to 2017, casinos operators persistent to introduce new table games to appeal gambler to enjoy total 27 table games in 2017. More than hundred types of online betting games are designed that many of them are similar to or improved of those table games.

**4.4 Industrial Policy and the evolution process of Atlantic City's gambling industry**

Atlantic City initial impacted gambling policy from 1969 when New Jersey voted to approve the lottery. The New Jersey lottery system began to operate in 1970, where pick3 was first national lottery game in 1975, followed by 7-11-21 lottery game in 1975, television lottery ticket and pick4 in 1977, which marked the beginning of introduction phase of ILC under the regulation of local government. In addition, referendum approved casino games within New Jersey in 1976, which the International Resort was the first casino in Atlantic City In 1978. A total of 12 casinos successively opened until the Ace Taj Mahal opened on April 4, 1990, which Atlantic City's gambling industry report a revenue surpassing Las Vegas and enter a development phase under legal supervision. Furthermore, Poker was introduced to Atlantic City in 1993 and Keno was introduced in 1994, New Jersey Lottery joined to the Intercontinental Lottery Association in 1999, Atlantic City's gambling industry entered a maturity phase with diversify regulation. Las Vegas style was introduced to Atlantic City when Borgata opened in 2003. While new gambling policies was authorized to the legalization gambling in the neighbor areas, Atlantic City's gambling industry entered a decline phase. When Sands closed in 2006 and following five casinos shut down, only seven casinos keep operating in Atlantic City. Until New Jersey licensed online gambling games and sports gambling in 2013, gambling industry went into a new period of recovery.

**4.5 Industrial innovation and the evolution process of Atlantic City's gambling industry**

Atlantic City explores changing the dilemma of gambling industry when it continually depress. Three casinos were authorized to operate 11 website platforms in 2013, and five casinos which Resorts, Caesars, Tropicana, Golden Nugget and Borgata are reported could operate online gambling on 22 website platforms in 2017. Arguably, online gambling has become an integral part of New Jersey's gambling industry when its revenue accounted for 4% of the state's total gambling revenue (TGR) in 2014, 6% of TGR in 2015, and 9.2% of TGR in 2017. After a decade of industry declines, internet gaming which plays innovation role to contribute a revenue growth respectively 32% in 2016 and 25% in 2017.

## 5 Discussion

Tourism is pillar industry for Atlantic City's in early 1920-1960, which was transformed into gambling industry in 1978 due to the recession of the tourism industry. After 40 years of the evolution of the gambling industry, Atlantic City gaming industry has entered the decline stage of ILC and formed a classic industrial life cycle course. It is proved that the gambling industry, as a special service industry, has a particular mechanism of ILC. The dynamic description of the life cycle of the Atlantic City gambling industry, which remaindered us a framework to uncover the decline of the Atlantic City gambling industry from the perspective of the characteristic variables in the life cycle phase of the gambling industry and obtain the relevant conclusions and enlightenment as following.

Gambling industry is the backbone of Atlantic City's economy affected by economic environment, it has been in decline since 2000 and exacerbated by the recession caused by financial crisis in 2008 within a global scope, while the city's local population fell to 35,000 with a 12 percentages drop from its peak. Considering its market conditions, Atlantic City fills with a saturated market with the total number of visitors growing rapidly from 7.08 million to 29,326 million between 1978 and 1985, reaching 3.4924 million in 2005. But the average annual visitor has been only 0.96 per cent growth rate over the past 20 years, and after 2005, it has decreased and the gambling industry has been relatively saturated. Furthermore, legalizing gambling around New Jersey areas forms a competitive environment, which opening of new casinos leads fierce competitive pattern in attracting gamblers from Atlantic City. While Pennsylvania is authorized five casino licenses in December 2006, attracting New York visitors to Connecticut's Revel Casino and Golden God Casino, which it's gambling industry quickly overtook New Jersey and become a second-highest-earning casino in the United States. The legalization of gambling in 24 states has further downplayed Atlantic City's gambling attraction.

Casino operators grew rapidly increase to 12 casinos after Atlantic City ligelitized gambling in 1978, which gambling revenues surpass Las Vegas in just 10 years. Companying newcomer's entering of industry, Atlantic City casinos quickly saturated at a peaking in 2006 where 12 casinos hair 35,000 employees within a small city who hosts only 35,000 residents. In addition, Atlantic City's gambling industry is highly monopolized, when the Trump family owning four casinos with 50 percent of the game tables and slot machines, which forms an oligopoly market that impedes market competition and industry entering of other enterprises, as well as new business ideas and product services into this area, only when Borgata brings the Lavages style to Atlantic City's gambling industry in 2002. As a result, industry innovation is hampered that competitive of gambling industry lad back within the surrounding area.

In addition to the above-mentioned gambling saturated market, the core benefits products of gambling have the most serious problem with the single gambling products in all casinos. Slot machine game dominate the market of core benefit gambling product, which accounts 70 percent of gambling revenue in Atlantic City where guests can't find any other recreational products besides gambling. Borgata's Las Vegas model brings lately some changes for Atlantic City's gambling product structure in 2002. Online gambling's entry changes the situation of Atlantic City gambling product structure in 2013, but new products which increase gradually market share is limited to the New Jersey market compare with other states. Furthermore, peripheral gambling products, comparing with core benefit gambling product, which are inadequate development such as catering, accommodation, shopping and other entertainment that results tourists an average of less than 24 hours staying in Atlantic City. The casino city named positioning of Atlantic City has hampered many years' development of other industries and unsuccessful rebuilt it like Las Vegas whom positioning is a diversified city converging with gambling entertainment and vacation tourism.

Atlantic City ignore the tourist preference changes in core benefit product. Traditional gambling industry dominated by core benefit gambling products for 40 years such as slot machines and table games though Atlantic City has transitioned from tourism to gambling since 1978. Casino revenues have increased, but the gambling industry has been losing money for years. Even in slot machine games, products with high bets of more than $5 are rarely on watched. The emergence of modern betting products VGTs or VLTs, as well as the emergence of online, sports and competitive betting is an aspect of the evolution of gamblers' preferences. Furthermore, the gambling peripheral products continually attract visitors spending more on food and other entertainment than gambling. Atlantic City neglects the change of perception of tourist spending which is distributed far more on non-gambling rather gambling.

Atlantic City gambling industrial policy adjustment lags behind the industry development, which industry policy focused on gambling industry development since legalized operation of gambling in 1976. Even gambling industry has entered a period of stability since 1990 that no corresponding industrial development strategy corrects during such a long period. Industrial development policies lag behind those gambling industries development in global and U.S., only passively following their market behavior and resulting a severe recession in the gambling industry. New industrial policies were introduced in 2012 such as sports betting and online gambling revise developing race in New Jersey. Although total gambling revenues began to rebound in 2016, the market remains a limited space. In addition, there is a serious squeeze on tourism and other industries, insufficient development of other local industries which attract fewer residents limiting the city space expanding with no sustainable development in the labor force and sustainable development.

**6 Conclusion**

The purpose of this paper was to examine whether industry variables actually exhibit a framework as hypothesized by ILC theory. Evidence which Atlantic City gambling ILC evolution reveals that dynamic evolution affected on microeconomic factors and its external the macroeconomic, market and demand characteristics. In addition, it validity proved by Atlantic City gambling ILC which give a clue to employ multi-indicators describing an industry evolution via longitudinal industry data interrelated in different stages of ILC.

Therefore, the study methodology applied to gambling industry allows us to confirm the validity and relevance of the ILC model as a tool to help practitioners to understand their industrial situation. The analysis has a number of caveats for operators and administrators that need to pay more attention to the development of the gambling industry with a accurately diagnose to the ILC phase which reginal gambling industry emerging and ongoing evolution, to adopt rational strategies to ensure its stable and sustainable development. According to the development mechanism of ILC, governments and casinos operators need to on watch the trend of market evolution, adopt market planning to maintain growth and prolong this stage. Especially when the gambling industry enters the mature stage, it is necessary to adopt more active strategies to prevent the quickly recession of industry. In addition, the gambling industry should attentive on watch the impacts from technological innovation on the gambling industry, especially network and communication technology currently bring a significant impacts and changes on gamblers' consumption behavior, to adjust their gambling products development strategy to prolong the stable phase and avoid the arrival of the recession of the gambling industry.